\documentstyle[aps,multicol]{revtex}
\begin{document}
\title{Electromagnetic Force in Dispersive and Transparent Media} 
\author{Yimin Jiang\\China Institute of Atomic Energy, 102413 Beijing, P. 
R. China\\
and\\ Mario Liu\\Institut f\"ur Theoretische Physik, Universit\"at Hannover,
\\30167 Hannover, Germany}
\date{\today}
\maketitle
\begin{abstract}
A hydrodynamic-type, macroscopic theory was set up recently to 
simultaneously account for dissipation and dispersion of electromagnetic 
field, in nonstationary condensed systems of nonlinear constitutive 
relations~\cite{JL}. Since it was published in the letter format, some 
algebra and the more subtle reasonings had to be left out. Two of the 
missing parts are presented in this paper: How algebraically the new results 
reduce to the known old ones; and more thoughts on the range of validity of 
the new theory, especially concerning the treatment of dissipation. 
\end{abstract}
\draft

\begin{multicols}{2}
\section{Introduction}

The macroscopic Maxwell equations, given in terms of  $\bf E,\ D,\ H$, and 
$\bf B$,  need constitutive relations linking them to be closed. The 
form of these relations depend crucially on two physical parameters, the 
frequency and the field strength. Weak fields are necessary for the validity 
of the linear response theory, $\bf E\sim D, H\sim B$; while the 
hydrodynamic constitutive relations~\cite{nurmax} presuppose small 
frequencies. 

While the limits on the constitutive relations are well known, the equally 
important closure problem one level higher has not entered the general 
awareness: A electromagnetic theory is only complete if it also considers 
the feedback, the force on the volume elements of condensed matter, exerted 
by the electromagnetic field. In the microscopic electrodynamics, this is 
simply the Lorentz force -- the Maxwell equations account for the field 
produced by the sources, while the Lorentz force (in conjunction with the 
Newton equation) describes how the field changes the position and the motion 
of the sources. In the macroscopic theory, two quantities are necessary to 
close the theory: (i) The additional energy due to the presence of the 
electromagnetic field, and (ii) the flux of the conserved, total momentum 
density, ie the total stress tensor including both the material and the 
field contribution. The hydrodynamic theory provides unambiguous expressions 
for both, and is therefore closed and complete. Circumstances are less 
fortunate for the linear response theory, as these expressions are only 
known with a string of additional restrictions. 

Assuming transparency (ie lack of dissipation), quasi-monochromatic 
external field and stationarity (ie identically vanishing velocity field of 
the condensed system) --- all in addition to the linearity of the 
constitutive relations --- Brillouin obtained the field energy in 1921, 
while Pitaevskii, forty years later, arrive at the attendant expression for 
the total stress tensor, cf \S80, 81 of the classic book of Landau and 
Lifshitz~\cite{LL8}, and the comprehensive and informative revew article 
by Kentwell and Jones~\cite{kenjo}. 

If we draw a diagram of field strength versus frequency $\omega$, with the 
field strength pointing to the right, and $\omega$ upward, a vertical stripe 
along the $\omega$-axis represents the range of validity for the linear 
response theory, while the hydrodynamic theory reigns within a horizontal 
stripe along the field axis. The expressions of Brillouin and Pitaevskii are 
valid in isolated patches in the vertical stripe, wherever dissipation is 
negligible~\cite{patch}. 

The parameter space beyond the above two perpendicular stripes needs a 
theory that can simultaneously account for dissipation, dispersion, 
nonlinear constitutive relations and finite velocities. Although one might 
expect principal difficulties in setting up such a theory, as neither of the 
two parameters, field and frequency, remains small, we are (up to and maybe 
slightly beyond the optical frequencies $\approx10^{15}$Hz) still in the 
realm of macroscopic physics,  as the electromagnetic wavelength remains 
large compared to the atomic graininess. And when asking questions such as 
what is the force on a volume element exerted by a strong laser beam, if we 
confine our curiosity to the averaged force --- with a temporal resolutions 
larger than the time needed to establish local equilibrium --- a simple, 
universal and hydrodynamic-type theory is still possible. Such a theory was 
derived recently, which includes the dynamics of polarization, but (in this 
first step) neglects magnetization~\cite{JL}. 

In this paper, we show explicitly that the new and nonlinear expressions for 
the energy and the stress tensor indeed reduce to the known ones, of Brillouin 
and Pitaevskii, in the specified limit. Because the nonlinear theory is the result 
of a qualitatively different, hydrodynamic-type approach, and not simply 
a ``generalization" of the linear theory, this outcome is by no mean obvious and assured. 
Besides, the associated algebra is fairly involved and needs to be 
presented. Once accomplished, this provides two additional bonuses: It  
strengthens our trust in the new theory and provides a transparent interpretation 
for the old and classic results which, having relied heavily on algebra, are 
somewhat lacking in appropriate physical pictures. 

The hydrodynamic theory of dispersion will be presented in section 
\ref{el}, to render the present manuscript self-contained. Because we shall 
only list the relevant formulas and abstain from repeating all the reasoning 
and arguments that lead to the new theory, the reader is advised to 
also read~\cite{JL}. However, the important question on the range of 
validity of the new theory in discussed here in greater details than 
in~\cite{JL}, at the beginning of the next section. Section \ref{mono} 
incorporates the specified approximation and deduce four results. They are 
compared to the energy density by Brillouin, and to three formulas by 
Pitaevskii's: the total stress tensor, the ``nonmagnetic" magnetization, and 
the time dependent permittivity. Section \ref{dis} ends with a brief 
summary. 

\section{The Hydrodynamic Theory of Dispersion}\label{el}
In this section, we shall first discuss in some details the range of 
validity of the new theory, then present its complete set of equations, and 
specify the theory by an expansion of the thermodynamic energy to third 
order in the field variables. 

\subsection{The Range of Validity\label{rang}}
The proper hydrodynamic theory of electromagnetism \cite{nurmax} accounts 
for the macroscopic dynamics of continuous media in the low frequency limit, 
for a system that is charged or exposed to external fields. Local 
thermodynamic equilibrium holds, and the set of hydrodynamic variables is 
identical to that of the thermodynamic variables. The equations of motion 
are conservation laws and the Maxwell equations, including irreversible 
terms accounting for dissipation.

At higher frequencies, microscopic variables deviate more and more from 
equilibrium, becoming independent, and finally ballistic. Denoting the time 
$\tau_{loc}$ needed to establish local equilibrium, this starts to happen 
when $\omega\tau_{loc}$ is no longer small.  To account for this 
circumstance, we usually have to abandon the hydrodynamic theory and embrace 
the Boltzmann theory which, despite its undeniable usefulness, is both a far 
more complex and a less general theory -- it considers the vast number of 
microscopic degrees of freedom explicitly, and it is confined to dilute 
systems.  The question therefore is whether an appropriately generalized 
hydrodynamic theory can be made to account for some of the more interesting 
aspects at higher frequencies, and save us from the complexities of the 
Boltzmann theory.  

Let us concentrate on one specific microscopic variable, the polarization 
${\bf P}$. Actually, as far as its spatial extent is concerned, it is a 
macroscopic rather then a microscopic variable, but it is certainly 
dependent in the hydrodynamic limit, as long as $\omega\tau_P\ll1$, where 
$\tau_P$ is ${\bf P}$'s  longest time scale. (All this can also be said of 
the magnetization, which we however shall not consider here.) 

In a dielectric medium, ${\bf P}$ has many characteristic times, given by 
the resonance frequencies and their widths. If they are well separated, then 
the equation of motion, close to one resonance and in the simplest case 
considered below is  \begin{equation}\label{d6}
\ddot{\bf P}/\omega^2_p-\tau\dot{\bf P}+{\bf P}=\chi{\bf D},\end{equation}
cf Eq(\ref{2.7}) of the next section. The given resonance may be overdamped 
($\tau\gg1/\omega_p$) or sharply resonating ($\tau\ll1/\omega_p$), and the 
characteristic time $\tau_P$, after which the polarization is no longer 
independent, is respectively $\tau$ and $2/(\omega_p^2\tau)$, while 
$1/\omega_p$ is typical for the time scale of $\bf P$'s motion in the 
resonating case. (Note that going to a different resonance, $\tau_P$ will 
change, it therefore depends on the frequency of the external field.)

If the polarization ${\bf P}$ is a specially slow variable, 
$\tau_P\gg\tau_{loc}$, (where $\tau_{loc}$ is around $10^{\rm -10}$s at 
usual temperatures and densities,) we may increase the range of validity of 
the hydrodynamic electromagnetic theory, from $\omega\ll1/\tau_P$ to 
$\omega\ll1/\tau_{loc}$, by taking the energy as a local function also of 
${\bf P}$, $\dot{\bf P}$, and derive the equation of motion for ${\bf P}$. 
There are quite a number of systems with a large $\tau_P$: All 
electro-rheological fluids have especially large $\tau_P$, of the order of 
$10^{\rm -4}$s, but  other complex fluids with large molecules and a 
permanent molecular dipole moment (such as nematic liquid crystals) also 
have a slow polarization. (Even the comparatively small water molecule with 
its permanent dipole moment has a $\tau_P$ around $10^{\rm -9}$s, just 
slightly too fast.) The hydrodynamic theory of dispersion presented below is 
unqualifiedly valid for these systems, (though it should usually be enough 
to neglect $\ddot{\bf P}$ in the equation of motion, or equivalently, 
exclude  $\dot{\bf P}$ as an additional variable.) We shall refer to this 
scenario as the hydrodynamic dispersion. 

Interestingly, essentially the same set of equations also accounts for a 
system in the ballistic regime, for quickly oscillating electric fields and 
polarizations, $1/\omega_p\ll\tau_{loc}$ --- if we confine our curiosity to 
questions such as what is the averaged force that a high frequency external 
field exerts on a volume element. (Note that the low resolution is quite 
sufficient for resolving the hydrodynamic responses to a high frequency 
field.)  This scenario we shall refer to as ballistic dispersion. Clearly, 
we need to understand why the same set of equations also works for the 
ballistic dispersion, and what the restrictions are.

First, a coarse grained, hydrodynamic-type description is at all possible 
because the field variables $\bf P$, $\bf D$ and $\bf B$ vary (due to the 
largeness of the light velocity) on macroscopic, hydrodynamic length scales. 
Second,  most of the general principles used as input to consider 
hydrodynamic dispersion are also valid here. Especially, the total energy 
and momentum remain conserved. The one exception is local equilibrium, the 
lack of which introduces some caveats with respect to temporal resolution 
and to dissipation. More specifically, they are: 
\begin{itemize}
\item In the ballistic regime, the variables of the theory divides into two 
types, fast and slow. The field variables are fast, the rest is slow. They 
are the densities of mass $\rho$, entropy $s$, total energy $u$, and total 
momentum $\mbox{\boldmath$g$}^{\rm tot}$. Therefore, the equations of motion 
of the field variables are highly accurate, resolving temporal increments 
much less than $1/\omega_p$, while the actual hydrodynamic equations are 
much coarser, with a resolution low compared to $\tau_{loc}$. And because 
every differential equation, consistently, has a unique resolution, all 
field terms appearing in the slow, hydrodynamic equations need to be 
appropriately averaged. 
\item The generalized hydrodynamic theory presented 
in the next section is shown below to be clearly valid for ballistic 
dispersion in the transparent region, where electromagnetic dissipation is 
negligible, and the time $\tau_P$ with which $\bf P$ looses energy long. But 
as argued below, it should remain valid even if field dissipation is strong. 
\end{itemize}

Taking electromagnetic dissipation into account, the total, conserved 
energy divides into three parts, 
\begin{equation}\label{1.1}
U=U^{\rm mat}+U^{\rm em}+U^{\rm mic}. \end{equation}
The first is the thermodynamic energy in the absence of an external field; 
the second is the additional energy in the presence of a field; and the 
third is the rest, the energy of all microscopic variables not given in the 
first two explicitly, $U^{\rm mic}(x_1^2, x_2^2\dots)$. The variables $x_i$ 
are defined such that they vanish in local equilibrium, so they are 
irrelevant for the consideration of hydrodynamic dispersion.  In the 
ballistic regime, $U^{\rm mic}$ is finite and serves as a transit hall: 
External energy is being fed continually into $U^{\rm em}$, the 
electromagnetic dissipation excite some microscopic degrees of freedom 
$x_i$, and convert $U^{\rm em}$ into $U^{\rm mic}$ --- which after the 
comparably long time of $\tau_{loc}$ becomes heat, $U^{\rm em}\to U^{\rm 
mic}\to\int T{\rm d}s$.  The rate at which $U^{\rm em}$ is lost is 
approximately $\dot U^{\rm em}\approx-U^{\rm em}/\tau_P$, the average time 
this energy stays in the transit hall is $\tau_{loc}$, so $U^{\rm 
mic}\approx (\tau_{loc}/\tau_P)U^{\rm em}$. The right side translates into 
$(\tau_{loc}/\tau)U^{\rm em}$ for the overdamped oscillation, and into 
$\textstyle\frac{1}{2}(\omega^2_p\tau_{loc}\tau)U^{\rm em}$ for the 
resonating one. In the first case, we always have $U^{\rm mic}\gg U^{\rm 
em}$, and in the second we mostly do, rendering the transit hall usually 
large. 

Including nonhydrodynamic variables such as $\bf P$ leads to contributions 
$\sim{\bf P}$ and $\partial U/\partial{\bf P}$ in the energy and momentum 
flux, see next section. If $U^{\rm mic}$ is nonzero, we would expect similar 
terms $\sim x_i$ and $\partial U/\partial x_i$. These we may neglect in the 
transparent region of vanishing dissipation, defined as the frequency regime 
where the stringent condition $\omega^2_p\tau_{loc} \tau\ll1$ holds, or 
equivalently $U^{\rm mic}\ll U^{\rm em}$, so $U^{\rm mic}$ and its 
contributions may be neglected. Outside these regimes, circumstances are not 
as certain and in need of a clarifying, more microscopic approach such as 
the  Boltzmann theory, to confirm the considerations in the next paragraph. 

There are reasons why we may quite generally neglect $U^{\rm mic}$: While 
terms such as ${\bf P}^2$ and $\dot{\bf P}^2$, of macroscopic extent, 
coherently add up over many periods to yield diverse,  slowly varying 
contributions $\sim\langle{\bf P}^2\rangle$ and $\langle\dot{\bf 
P}^2\rangle$ in the momentum and energy flux, and thereby directly alter the 
slow, hydrodynamic variables, the quantities $x_i$ are random and of 
microscopic spatial scales. They further dissipate and degrade, to 
eventually turn into heat. So on a coarse, hydrodynamic time scale, we may 
simply lump $\langle U^{\rm mic}\rangle$ into heat $\int T{\rm d}s$, and 
$\langle\frac{\partial}{\partial t}U^{\rm mic}\rangle$ into the heat 
production $R$. Then, clearly, $U^{\rm mic}$ is neglected as an independent 
entity. 

On a more fundamental level, the very criterion by which we have singled 
out ${\bf P}$ and ${\bf\dot P}$ from the lot of microscopic degrees of 
freedom is because they are qualitatively different: Given a certain energy 
content in the field ${\bf D}$ and ${\bf B}$, there is a back and forth of 
energy flow between ${\bf D}$, ${\bf B}$, ${\bf P}$ and ${\bf\dot P}$  
within each period; while the field energy that leaks into the other 
microscopic degrees of freedom is usually lost. In fact, for an overdamped 
resonance, it is (as mentioned) appropriate to exclude ${\bf\dot P}$ as an 
explicit variable, and consider it as one of the many ordinary microscopic 
degrees of freedom, as the energy leaked into ${\bf\dot P}$  is lost to the 
field. On the other hand, if a system involves more variables in the 
tidal-like transfer of field energy, the present theory needs to be 
generalized to also include them --- one example comes readily to mind: an 
independent magnetization. 

Note that being a function also of the slow variables, $U^{\rm em}$ is not 
conserved by itself, and the permeability $\varepsilon$ will in general 
contain an imaginary part to account for this fact, even without any 
dissipation (or electric charge).    

\subsection{The Equations of Motion\label{eom}}
The complete hydrodynamic theory of dispersion consists of a closed set of 
partial differential equations that governs the dynamics of the medium and 
the electromagnetic field. The structure of the equations is determined by 
general principles: the Maxwell equations, the Lorentz-Galilean 
transformation, the thermodynamic theory and the relevant conservation laws. 

Combining the two macroscopic energy densities 
\begin{equation}\label{1.2}
U^{\rm Mac}\equiv U^{\rm mat}+U^{\rm em},
\end{equation} 
we take it as a function of the entropy density $s$, mass density $\rho$, 
the electric and magnetic field {\bf D} ~and {\bf B}, the electric 
polarization {\bf P}, its canonical conjugate {\bf a} (that will turn out to 
be essentially $\sim\bf\dot P$), and the thermodynamic momentum density 
\mbox{\boldmath$g$}, \begin{eqnarray}
{\rm d}U^{\rm Mac} = T{\rm d}s+ \mu {\rm d}\rho+ \mbox{\boldmath$v$}\cdot 
{\rm d}\mbox{\boldmath$g$}+ {\bf E}\cdot {\rm d}{\bf D}\nonumber\\
+{\bf H}\cdot {\rm d}{\bf B} +{\bf h}\cdot {\rm d}{\bf P} +{\bf b}\cdot {\rm 
d}{\bf a}, \label{2.2}
\end{eqnarray}
where the thermodynamic momentum density \mbox{\boldmath$g$} is related to 
the total momentum density \begin{equation}
\mbox{\boldmath$g$}^{\rm tot}= \rho\mbox{\boldmath$v$} +({\bf E}\times{\bf 
H})/c
\end{equation}
through \cite{{henjes},{HL}}
\begin{equation}
\mbox{\boldmath$g$}=\mbox{\boldmath$g$}^{\rm tot}-{\bf D}\times{\bf B}/c. 
\label{2.1}\end{equation}
As soon as the energy function $U^{\rm Mac}$ is known, the temperature $T$, 
chemical potential $\mu$, velocity \mbox{\boldmath$v$}, field strengths {\bf 
E} ~and {\bf H} ~are also determined.  (In accordance with \cite{JL}, the 
polarization defined here is a rest frame quantity, ${\bf P}\equiv{\bf 
D}_0-{\bf E}_0$.) 

Isotropy of space results in the identity
\begin{equation}
{\bf E}\times{\bf D}+{\bf H}\times{\bf B}+{\bf h}\times{\bf P}+{\bf 
b}\times{\bf 
a}+\mbox{\boldmath$v$}\times\mbox{\boldmath$g$}=0.\label{2.3}\end{equation}
The Maxwell equations 
\begin{eqnarray}
\nabla{\bf B}&=&0,\quad{\dot{\bf B}}=-c\nabla\times{\bf E},\nonumber\\
\nabla{\bf D}&=&\rho^e,\quad{\dot{\bf D}}=c\nabla\times{\bf 
H}-\rho^e\mbox{\boldmath$v$}
\label{2.4}\end{eqnarray}
account for the motion of {\bf D}~and {\bf B}. Here, the dot indicates 
partial temporal derivative $\partial/\partial t$ and $\rho^e$ denotes the 
macroscopic charge density.  The variables $\rho$, $U$, 
$\mbox{\boldmath$g$}^{\rm tot}$ are conserved, their equations of motion 
take the form

\begin{eqnarray}
{\dot\rho}+\nabla\cdot(\rho\mbox{\boldmath$v$})=0,\label{2.5}\\
\dot U+\nabla\cdot{\bf Q}=0,\label{2.5c}\\
{\dot g_i}^{\rm tot}+\nabla_j(\Pi_{ij}-\Pi^D_{ij})=0, 
\label{2.5a}\end{eqnarray}
where {\bf Q}~is the total energy flux, and $(\Pi_{ij}-\Pi^D_{ij})$ the 
symmetric total momentum flux, or total stress tensor. The entropy is not 
conserved, and has a positive source $R$, 

\begin{equation}
\dot s+\nabla\cdot(s\mbox{\boldmath$v$}-{\bf f}^D)=R/T,\quad R\geq 0,
\label{2.5b}\end{equation}
The dissipative part of entropy flux $f^D$ describes especially 
thermoconduction, while $\Pi^D_{ij}$ accounts primarily for 
viscosity-related phenomena.

The macroscopic variables {\bf P} and {\bf a} are governed by equations 
that are essentially of the Hamiltonian type, $${\dot {\bf P}}=\partial 
U/\partial {\bf a}={\bf b},\quad{\dot {\bf a}}=-\partial U/\partial {\bf 
P}=-{\bf h},$$ with some supplementary terms needed to ensure the proper 
transformation behavior, and to account for dissipation. First, the temporal 
derivative is replaced by the Galilean invariant operator that takes into 
account the effect of the local movement of the medium, \begin{equation}
D_t=\partial_t+(\mbox{\boldmath$v$}\nabla)- \Omega\times,
\label{2.6}\end{equation}
where $\Omega\equiv{\textstyle\frac{1}{2}}\nabla\times\mbox{\boldmath$v$}$. 
Second, a dissipative force ${\bf h}^D$ is introduced in the equation for 
{\bf a} to account for electromagnetic dissipation that (in the linear case) 
is usually taken care of by an imaginary term in the electric permittivity 
$\varepsilon$. Third, the polarization is changed if the medium undergoes 
volume dilatation, as a term $-{\bf P}(\nabla\mbox{\boldmath$v$})$ appears 
in the equation of motion for {\bf P}, \begin{equation}
D_t{\bf P}={\bf b}-{\bf P}(\nabla\cdot\mbox{\boldmath$v$}),\quad D_t{\bf 
a}=-{\bf h}-{\bf h}^D.
\label{2.7}
\end{equation}

The dissipative terms are determined with the familiar method of 
irreversible thermodynamics: We first identify the entropy production as 
\begin{equation}
R+\dot U^{\rm mic}={\bf f}^D\cdot(\nabla T)+{\bf h}^D\cdot{\bf 
b}+\Pi^D_{ij}v_{ij},
\label{2.8}\end{equation}
where $v_{ij}={\textstyle{1\over2}}(\nabla_iv_j+\nabla_jv_i)$. Then take 
the fluxes as proportional to the thermodynamic forces,
\begin{equation}\label{2.9}
\left(\matrix{\Pi_{ik}^D\cr f_i^D\cr h_i^D\cr}\right)=
\left(\matrix {\eta_{ikjl}&\alpha_{ikj}&\beta_{ikj}\cr
{\bar\alpha}_{ijl}&\kappa_{ij}&\lambda_{ij}\cr
{\bar\beta}_{ijl}&{\bar\lambda}_{ij}&\zeta_{ij}\cr
}\right)\!\times\!
\left(\matrix{v_{jl}\cr \nabla_jT\cr b_j\cr}\right)
\end{equation}
(Appropriate Onsager reciprocity relations are implied.)

The energy flux is
\begin{eqnarray}\label{2.11a}
Q_i&=&(Ts+\mu\varrho+v_kg_k)v_i-Tf^D_i-v_j\Pi^D_{ji}+c{\bf (E\times 
H)_i}\nonumber\\ &+&v_i({\bf h}\cdot {\bf P}) 
+\textstyle\frac{1}{2}[\mbox{\boldmath$v$}\times ({\bf h}\times{\bf P} +{\bf 
b}\times{\bf a})]_i.
\end{eqnarray}
(The last two terms were erroneously omitted from \cite{JL}.) This 
expression may be rewritten as
\begin{eqnarray}
Q_i=c({\bf E}^0\!\times\!{\bf H}^0)_i-f_i^DT+U^{\rm Mac}v_i\nonumber\\ 
+(\Pi_{ij}-\Pi^D_{ij})v_j-v_kg_k^{\rm tot}v_i, \label{2.11}\end{eqnarray}
to see that the velocity-dependent terms do come from an Lorentz-Galilean 
transformation, discussed eg in \cite{J}. (${\bf E}^0\equiv{\bf E}+ 
\mbox{\boldmath$v$}\times{\bf B}/c$ and ${\bf H}^0\equiv{\bf 
H}-\mbox{\boldmath$v$}\times{\bf D}/c$ are the restframe fields.)  

The stress tensor is symmetric and given as 
\begin{eqnarray}
\Pi_{ij}&=&\textstyle\frac{1}{2}[v_ig_j-E_iD_j-H_iB_j+(i\leftrightarrow j)]+
(Ts+ \mu\rho\nonumber\\&+&\mbox{\boldmath$g$}\cdot\mbox{\boldmath$v$} +{\bf 
H}\cdot{\bf B}+ {\bf E}\cdot{\bf D}+ {\bf h}\cdot{\bf P}-U^{\rm 
Mac})\delta_{ij}.
\label{2.12}
\end{eqnarray}

Frequently, there are many different resonance frequencies of the 
polarization, not just the single one, given here as $\omega_p$. This fact 
can be accounted for by introducing as many ``subpolarizations'',
\begin{equation}
{\bf D-E=P}=\sum\ {\bf P_\alpha}, 
\label{c6}\end{equation}
chosen such that the two squared order terms of the energy are diagonal,
\begin{equation}
U^{\rm em}_0=\dots+\sum({\bf P_\alpha}^2/\chi_\alpha 
+\chi_\alpha\omega_\alpha^2{\bf a_\alpha}^2)/2 +\dots 
\label{c7}\end{equation}
Close to one resonance $\alpha$, if it is well separated, as all the other 
subpolarizations are not excited, we may simply substitute  ${\bf P}_\alpha$ 
for ${\bf P}$. 

\subsection{Some Explicit Expressions\label{expl}}
Now, the above equations are rendered more explicit by an expansion of the 
energy function
in the vector-variables {\bf D}, {\bf B}, {\bf P}, {\bf a} and 
\mbox{\boldmath$v$} to third order, as this is sufficient for a comparison 
to the linear results by Brillouin and Pitaevskii's. For a magnetically 
inactive medium (ie taking the static magnetic permeability as 1), such an 
expansion yields
\begin{eqnarray}
U^{\rm Mac}&=&U^{\rm 
mat}+\textstyle\frac{1}{2}B^2+\textstyle\frac{1}{2}D^2-{\bf D}\cdot{\bf P}+
\textstyle\frac{1}{2}P^2/\chi 
\nonumber\\&+&\textstyle\frac{1}{2}\chi\omega_p^2a^2
-\xi{\bf B}\cdot({\bf P}\times{\bf a}) +\textstyle\frac{1}{2}\rho v^2+{\cal 
O}^4,
\label{2.13}\end{eqnarray}
where ${\cal O}^n$ denotes terms of n-th or higher order in the
vector variables $({\bf D},{\bf B},{\bf P},{\bf a},\mbox{\boldmath$v$})$.
The energy density in the absence of electromagnetic fields is $U^{\rm 
mat}(s,\rho)$; the coefficient ${\chi}$ is related to the conventional 
static dielectric susceptibility $\chi'=P/E$ by  ${\chi'}^{\rm 
-1}={\chi}^{\rm -1}-1$;
${\omega_p^2}$ is the dielectric resonance frequency; $\xi$ is connected to 
the magnetic cyclotron-frequency $\omega_B$; all these parameters are in 
principle functions of $\rho$ and $s$. 

Obtaining the differential form from Eqs(\ref{2.13}, \ref{2.1}) 
$$
d(U^{\rm Mac}-\mbox{\boldmath$v$}\cdot\mbox{\boldmath$g$})=Tds+\mu 
d\rho+{\bf E}\cdot d{\bf D} +{\bf H}\cdot d{\bf B}
$$
\begin{equation}
+{\bf h}\cdot d{\bf P} +{\bf b}\cdot d{\bf a} -\mbox{\boldmath$g$}\cdot 
d\mbox{\boldmath$v$},
\label{2.15}
\end{equation}
we can derive the thermodynamically conjugate variables,  
\begin{eqnarray}
T&=&{\partial U^{\rm mat}\over\partial s}+{P^2\over2}{\partial{\chi}^{\rm 
-1}\over\partial s}
+{a^2\over2}{\partial{\chi}{\omega_p^2}\over\partial s}\nonumber\\
&-&{\bf B}\cdot({\bf P}\times{\bf a}){\partial\xi\over\partial s}+{\cal 
O}^4,\label{2.15a}\\
\mu&=&{\partial U^{\rm mat}\over\partial 
\rho}+{P^2\over2}{\partial{\chi}^{\rm -1}\over\partial \rho }
+{a^2\over2}{\partial{\chi}{\omega_p^2}\over\partial \rho }\nonumber\\
&-&{\bf B}({\bf P}\times{\bf a}){\partial\xi\over\partial \rho 
}-{1\over2}v^2+{\cal O}^4,
\label{2.16}\\
{\bf E}&=&{\bf D}-{\bf P}-\mbox{\boldmath$v$}\times{\bf M}+{\cal O}^3, 
\label{2.14}\\
{\bf H}&=&{\bf B}-\xi({\bf P}\times{\bf a})+{1\over 
c}\mbox{\boldmath$v$}\times{\bf P}+{\cal O}^3\label{2.17}\\
{\bf h}&=&{1\over{\chi}}{\bf P}-{\bf D}-\xi{\bf a}\times{\bf B}+{1\over 
c}{\bf B}\times\mbox{\boldmath$v$}+{\cal O}^3,\label{2.18}\\
{\bf b}&=&{\chi}\omega_p^2{\bf a}-\xi{\bf B}\times{\bf P}+{\cal 
O}^3.\label{2.19}
\end{eqnarray}

Note that the magnetization ${\bf M}\equiv{\bf B}-{\bf H}$ as given in 
Eq(\ref{2.17}) is a term of order ${\cal O}^2$. So the difference in the 
polarization $\mbox{\boldmath$v$}\times{\bf M}$, between the rest frame 
quantity ${\bf D}_0-{\bf E}_0$ and the laboratory quantity ${\bf D}-{\bf 
E}$, is of order ${\cal O}^3$. Within the accuracy of the above equations, 
it is therefore ignored.

Inserting (\ref{2.19}) in (\ref{2.7}), we obtain the expression
\begin{equation}
{\bf a}={1\over{\chi}\omega_p^2}[D_t{\bf P}+{\bf 
P}(\nabla\mbox{\boldmath$v$})+\xi{\bf B}\times{\bf P}]+{\cal O}^3
\label{2.20}
\end{equation}
that may be used to eliminate {\bf a}~ in the above formulas, and write 
them instead with
${\bf\dot P}$. Especially, the magnetization in the rest frame and to 
lowest order is then $(\xi/{\chi}{\omega_p^2}){\bf P}\times{\dot{\bf P}}$. 
The nonmagnetic magnetization is now seen to result from rotations of the 
polarization. 

For a qualitative estimate of the coefficient $\xi$, envision electrons 
revolving around their ion centers \cite{J}. Assuming the rotations to occur 
in phase, the magnetization associated with it is $(q_en_e/2m_e){\bf L}$, 
where $q_e$ and $m_e$ are the charge and mass of the electrons, while $n_e$ 
denotes their density.The angular moment of the electrons, ${\bf L} =m_e{\bf 
r}_e\times{\dot{\bf r}}_e$ (with ${\bf r}_e$ the radius of the circular 
motion) can also be written as ${\bf L}=(m_e/q_e^2{\omega_p^2}){\bf 
P}\times{\dot{\bf P}}$,  because the polarization {\bf P}~ is $q_en_e{\bf 
r}_e$. The attendant magnetization is ${\bf M}=(1/2q_en_e){\bf P}\times 
{\dot{\bf P}}$. So the coefficient $\xi$ is
\begin{equation}\label{2.22}
\xi=\chi\omega^2_p/2q_en_e.
\end{equation}
Particularly for an electron plasma, $\chi=1$, and $\omega_p$ can be
considered as the plasma frequency $(q_e^2n_e/m_e)^{1/2}$.
Eq(\ref{2.22}) reduces to $\xi=q_e/2m_e=-\omega_B/2B$, with
$\omega_B=-Bq_e/m_e$ the plasma cyclotron frequency.

\section{Monochromatic approximations}\label{mono}

With the help of a closed L-C circuit, Pitaevski${\breve{\rm i}}$ obtained 
a number of
important results on the effects of a high-frequency field in a medium 
\cite{pita}.
Because of the special setup, the results are subject to certain 
restrictions. In order to
compare our theory with his work, the same limits will be taken in our 
theory. Therefore, we shall consider a transparent medium exposed to a 
strictly monochromatic electric field:
\begin{equation}
{\bf E}={1\over2}{\cal E} e^{\rm -i\omega t}+c.c.,\qquad{\dot{\cal E}}=0,
\label{3.1}
\end{equation}
where ${\cal E}$ is the constant amplitude and $\omega$ the frequency. From 
now on, we shall  always assume that the medium is at rest, so any 
velocity-dependent terms will be discarded. Because only the electric 
properties are of interest, we also omit the quickly oscillating part of 
magnetic field in the medium, as in \cite{pita}, though a strong, constant 
magnetic field is allowed to be present. If the material coefficients 
${\chi}$, ${\omega_p^2}$, $\xi$ are constant with respect to time, the 
induction {\bf D}~ and polarization {\bf P}~ will also take the 
monochromatic form
\begin{equation}
{\bf D}={1\over2}{\cal D} e^{\rm -i\omega t}+c.c.,~
{\bf P}={1\over2}{\cal P} e^{\rm -i\omega t}+c.c..
\label{3.2}
\end{equation}
However, if ${\chi}$, ${\omega_p^2}$, $\xi$ and {\bf B}~are allowed to vary 
-- slowly -- via their dependence on the density or temperature, the fields 
{\bf D} and {\bf P}~will become quasi-monochromatic. In this case, we may 
still hold $\omega$ to be strictly constant, while allowing the amplitudes 
${\cal D}$, ${\cal P}$ to chang slowly with time. The quasi-monochromatic 
situation will be studied in subsection \ref{time}. 

In what follows, the dynamic equations given in the previous section will 
be investigated, under the preconditions mentioned above. We will show in 
detail the derivations of four formulas, all well-known in the 
literature~\cite{kenjo}.

\subsection{The permittivity\label{perm}}
The frequency-dependent permittivity $\varepsilon_{ij}$ is calculated from 
the equation of motion for {\bf P}. The expression is given by inserting 
(\ref{2.20}) and (\ref{2.18}) to (\ref{2.7}), taking  the coefficients 
${\chi}$, $\xi$ and the magnetic field {\bf B} as constants. Neglecting the 
velocity-dependent terms, we have \begin{equation}
{\bf D}-{{\bf P}\over{\chi}}-2{\xi{\bf B}\times{\dot{\bf 
P}}\over{\chi}\omega_p^2}-
{{\ddot{\bf P}}\over{\chi}\omega_p^2}=0.
\label{3.3}\end{equation}
If the fields {\bf E}, {\bf P} assume the monochromatic form of 
Eqs(\ref{3.1},\ref{3.2}), the above equation becomes
\begin{equation}
\left(1-{1\over{\chi}}+{\omega^2\over{\chi}{\omega_p^2}}\right){\cal P}
+2i{\xi\omega\over{\chi}{\omega_p^2}}{\bf B}\times{\cal P}+{\cal E}=0.
\label{3.4}
\end{equation}
Solving it for ${\cal P}$, we obtain
\begin{equation}
{\cal P}=(\varepsilon_1-1){\cal E}+\varepsilon_2({\bf B}{\cal E}){\bf 
B}+i\varepsilon_3({\cal E}\times{\bf B}),
\label{3.5}\end{equation}
with
\begin{equation}
\varepsilon_1=1-{{\chi}{\omega_p^2}(\omega^2-{\omega_p^2}+ 
{\chi}{\omega_p^2})\over(\omega^2-{\omega_p^2}+{\chi}{\omega_p^2})^2- 
\omega^2\omega_B^2},
\label{3.6}
\end{equation}
\begin{equation}
\varepsilon_2={{\chi}{\omega_p^2}\omega^2\omega_B^2\over
[(\omega^2-{\omega_p^2}+{\chi}{\omega_p^2})^2-\omega^2\omega_B^2]
(\omega^2-{\omega_p^2}+{\chi}{\omega_p^2})B^2}
\label{3.7}
\end{equation}
and
\begin{equation}
\varepsilon_3=
{{\chi}{\omega_p^2}\omega\omega_B\over[(\omega^2-{\omega_p^2}+ 
{\chi}{\omega_p^2})^2-\omega^2\omega_B^2]B},
\label{3.8}
\end{equation}
where 
\begin{equation}
\omega_B=-2\xi B.
\label{3.9}
\end{equation}

Using the fact that ${\cal P}_i= (\varepsilon_{ik}-\delta_{ik}){\cal E}_k$, 
we observe from (\ref{3.5}) that the permittivity is
\begin{equation}
\varepsilon_{ik}=\varepsilon_1\delta_{ik}+\varepsilon_2B_iB_k+ 
i\varepsilon_3\epsilon_{ikl}B_l,
\label{3.10}
\end{equation}
 where $\epsilon_{ikl}$ is the total antisymmetric tensor, 
$\epsilon_{123}=1$.
In the low frequency limit $\omega\to 0$, we have according to (\ref{3.6})
 $\varepsilon_1=1/(1-{\chi})$. Note that the imaginary term in (\ref{3.10}) 
is not connected to dissipation. It is a purely reactive term. This can best 
be see from its invariance under the time-reversal operation: $\omega\to 
-\omega$, $B_i\to -B_i$.

\subsection{The energy density\label{dens}}

 Eliminating the quantity {\bf a} in the energy function (\ref{2.13}) with 
the help of (\ref{2.20}), we get, for a medium at rest and including terms 
of third order in the field
\begin{equation}
 U=U^{\rm mat}+{B^2\over2}+{D^2\over2}-{\bf D}{\bf 
P}+{P^2\over2{\chi}}+{{\dot P}^2\over2{\chi}{\omega_p^2}}.
\label{3.11}
\end{equation}
 Now consider the monochromatic case (\ref{3.1},\ref{3.2}) and apply a  
time-average procedure denoted as ${\langle \cdots \rangle }$, the energy 
density is then given as $$\langle U\rangle =U^{\rm 
mat}+{B^2\over2}+{1\over4}{\cal E}_k{\cal E}_k^*+{1\over4{\chi}} 
\left(1+{\omega^2\over{\omega_p^2}}-{\chi}\right){\cal P}_k{\cal P}_k^*$$
or
$$\langle U\rangle =U^{\rm mat}+{B^2\over2}+{1\over4}|{\cal E}|^2$$
\begin{equation}
+{1\over4}\left({1\over{\chi}}+{\omega^2\over{\chi}{\omega_p^2}}-1\right)
(\varepsilon_{km}-\delta_{km})
(\varepsilon_{kn}^*-\delta_{kn})
{\cal E}_m{\cal E}^*_n,
\label{3.12}\end{equation}
 where $^*$ denotes complex conjugation. Because in our work the 
permittivity is given by (\ref{3.10}) and (\ref{3.6}-\ref{3.8}), we can 
verify by direct computations that the equation
$$
{\partial\omega\varepsilon_{mn}\over\partial\omega}
=\delta_{mn}
+{1\over\chi}\left(1+{\omega^2\over{\omega_p^2}}-\chi\right)
(\varepsilon_{km}-\delta_{km})
$$
\begin{equation}
\times(\varepsilon_{kn}^*-\delta_{kn})
\label{3.13}
\end{equation}
 holds for this form of permittivity. So the time-averaged energy density 
could be expressed as
\begin{equation}
\langle U\rangle =U^{\rm mat}+{B^2\over2}+{1\over4}{\partial\omega 
\varepsilon_{mn}\over\partial\omega}
{\cal E}_m{\cal E}^*_n,
\label{3.14}
\end{equation}
 which is the Brillouin's expression for the time-averaged energy density of 
the electric field \cite{LL8}.

\subsection{Pitaevski${\breve{\rm i}}$ magnetization}
 The equation (\ref{2.17}) shows that a magnetization could be induced 
dynamically in an electrically polarizable medium, although the static 
magnetic permeability is 1. Inserting (\ref{2.20}) in (\ref{2.17}) we 
obtain, for a medium at rest,  the magnetization

\begin{equation}
{\bf M}={\xi\over{\chi}{\omega_p^2}}{\bf P}\times{\dot{\bf P}}+{\cal O}^3.
\label{3.15}
\end{equation}
 In the monochromatic approximation, the time-average of this magnetization 
is the same as that obtained by Pitaevski${\breve{\rm i}}$ \cite{pita}. 
Indeed, inserting
(\ref{3.1}, \ref{3.2}) in (\ref{3.15}) we have 

$$
 \langle {\bf M}\rangle _i={i\omega\xi\over2{\chi}{\omega_p^2}}({\cal 
P}\times{\cal P}^*)_i
$$
\begin{equation}
={i\omega\xi\over2{\chi}{\omega_p^2}}\epsilon_{ijk}(\varepsilon_{jm}- 
\delta_{jm})(\varepsilon_{kn}^*-\delta_{kn}){\cal E}_m{\cal E}_n^*.
\label{3.16}
\end{equation}
 With the help of the expression of the permittivity (\ref{3.10}), 
(\ref{3.6}-\ref{3.8}),
one can show the validity
of the equation
\begin{equation}
{i\omega\xi\over{\chi}{\omega_p^2}}\epsilon_{ijk}(\varepsilon_{jm}- 
\delta_{jm})
(\varepsilon_{kn}^*-\delta_{kn})={1\over2} 
{\partial\varepsilon_{nm}\over\partial B_i}.
\label{3.17}
\end{equation}
So (\ref{3.16}) possesses the form given by Pitaevski${\breve{\rm i}}$ 
\begin{equation}
\langle {\bf M}\rangle ={1\over4} {\partial\varepsilon_{mn}\over\partial{\bf 
B}}{\cal E}_m^*{\cal E}_n.
\label{3.18}
\end{equation}

\subsection{The stress tensor\label{stres}}
 Inserting the expressions (\ref{2.13}), (\ref{2.15}), (\ref{2.16}) for $U$, 
$T$, $\mu$
into the stress (\ref{2.12})
 and eliminating the fields {\bf B}, {\bf a}, {\bf h} with the help of 
(\ref{2.17}), (\ref{2.18}), (\ref{2.20}), the stress becomes

$$
\Pi_{ij}=
\Big[p_0+{H^2\over2}+{E^2\over2}+{1\over2}\left({1\over{\chi}}-1\right)P^2
-{{\dot P}^2\over 2{\chi}{\omega_p^2}}
$$
$$
-{\xi\over{\chi}{\omega_p^2}}{\bf H}({\bf P}\times{\dot{\bf P}})
+{P^2\over2}\left(\rho{\partial\over\partial\rho}{1\over{\chi}}\right)
+{1\over2}{{\dot P}^2\over {\chi}^2\omega_p^4}
\rho{\partial {\chi}{\omega_p^2}\over\partial\rho}
$$
$$
-{\bf H}({\bf P}\times{\dot{\bf P}})
\left(\rho{\partial\over\partial\rho}{\xi\over{\chi}{\omega_p^2}}\right)
\Big]\delta_{ij}
-H_iH_j
$$
\begin{equation}
-{1\over2}[E_iD_j+M_iH_j
+(i\leftrightarrow j)]+{\cal O}^4,
\label{3.19}
\end{equation}
 where $p_0$ is pressure of the medium in the case without electromagnetic 
fields,

\begin{equation}
 p_0(\rho,s)=-U^{\rm mat}+\rho{\partial U^{\rm 
mat}\over\partial\rho}+s{\partial U^{\rm mat}\over\partial s}.
\label{3.20}
\end{equation}
 Here, in order to avoid unnecessarily complicated formulas, we also neglect 
the entropy dependence of the parameters ${\chi}$, ${\omega_p^2}$, $\xi$  in 
(\ref{3.19}).
 When the two electric fields take the monochromatic form (\ref{3.1}, 
\ref{3.2}), we obtain after the time-averaging procedure, 

$$
\langle \Pi_{ij}\rangle =\Big\{
p_0+{H^2\over2}+{1\over4}{\cal E}{\cal E}^*-
 {\omega^2-{\omega_p^2}+{\chi}{\omega_p^2}\over4{\chi}{\omega_p^2}}{\cal 
P}{\cal P}^*
$$
$$
-{i\omega\xi\over2{\chi}{\omega_p^2}}({\cal P}\times{\cal P}^*){\bf H}
+{1\over4{\chi}^2{\omega_p^2} H}
$$
$$
\times
 \left[H(\omega^2-{\omega_p^2}){\cal P}{\cal P}^*-i\omega\omega_H ({\cal 
P}\times{\cal P}^*){\bf H}\right]
\rho{\partial{\chi}\over\partial\rho}
$$
$$
+{1\over4{\chi}\omega_p^4 H}
 \left[H\omega^2{\cal P}{\cal P}^*-i\omega\omega_H ({\cal P}\times{\cal 
P}^*){\bf H}\right]
\rho{\partial{\omega_p^2}\over\partial\rho}
$$
$$
+{i\omega\over4{\chi}{\omega_p^2} H} ({\cal P}\times{\cal P}^*){\bf H}
\rho{\partial\omega_H\over\partial\rho}
\Big\}\delta_{ij}-H_iH_j
$$
\begin{equation}
-{1\over8}({\cal E}_i{\cal D}_j^*+{\cal E}_j{\cal D}_i^*+c.c)
-{1\over2}(\langle M\rangle _iH_j+\langle M\rangle _jH_i).
\label{3.21}
\end{equation}
 Because the difference between {\bf B}~ and {\bf H}~ (ie the 
Pitaevski${\breve{\rm i}}$ magnetization (\ref{3.18})) is of
 second order in ${\cal E}$, we can write the formula (\ref{3.5}) for ${\cal 
P}$  to the same accuracy as

\begin{equation}
{\cal P}
 =(\varepsilon_1-1){\cal E}+\varepsilon_2({\bf H}{\cal E}){\bf 
H}+i\varepsilon_3({\cal E}\times{\bf H}),
\label{3.22}
\end{equation}
with

\begin{equation}
 \varepsilon_1=1-{{\chi}{\omega_p^2}(\omega^2-{\omega_p^2}+ 
{\chi}{\omega_p^2})\over(\omega^2-{\omega_p^2}+{\chi}{\omega_p^2})^2- 
\omega^2\omega_H^2},
\label{3.23}
\end{equation}
\begin{equation}
\varepsilon_2={{\chi}{\omega_p^2}\omega^2\omega_H^2\over
{[(\omega^2-{\omega_p^2}+{\chi}{\omega_p^2})^2-\omega^2\omega_H^2]}
(\omega^2-{\omega_p^2}+{\chi}{\omega_p^2})H^2}
\label{3.24}
\end{equation}
and

\begin{equation}
\varepsilon_3={{\chi}{\omega_p^2}\omega\omega_H\over[(\omega^2- 
{\omega_p^2}+{\chi}{\omega_p^2})^2-\omega^2\omega_H^2]H},
\label{3.25}
\end{equation}
where

\begin{equation}
\omega_H=-2\xi H.
\label{3.26}
\end{equation}
Now inserting in the stress (\ref{3.21}) the expression (\ref{3.22}), it 
becomes

$$
\langle \Pi_{ij}\rangle =\Big\{
p_0+{H^2\over2}-
{1\over4}\left(\rho{\partial\varepsilon_{lm}\over
\partial\rho}-\varepsilon_{lm}\right){\cal E}_l^*{\cal E}_m
\Big\}\delta_{ij}
$$
$$
-H_iH_j-{1\over8}({\cal E}_i{\cal D}_j^*+{\cal E}_j{\cal D}_i^*+c.c.)
$$
\begin{equation}
-{1\over2}(\langle M\rangle _iH_j+\langle M\rangle _jH_i),
\label{3.27}
\end{equation}
where we have used the following equations valid for the $\varepsilon_1$,
$\varepsilon_2$, $\varepsilon_3$ given by (\ref{3.23}-\ref{3.25}),
\begin{equation}
{\partial\varepsilon_1\over\partial{\chi}}=
{{\omega_p^2}-\omega^2\over{\chi}^2{\omega_p^2}}[(\varepsilon_1- 
1)^2+\varepsilon_3^2H^2]
-{2\omega\omega_HH\over{\chi}^2{\omega_p^2}}\varepsilon_3(\varepsilon_1-1),
\label{3.28}
\end{equation}
$$
{\partial\varepsilon_2\over\partial{\chi}}=
{{\omega_p^2}-\omega^2\over{\chi}^2{\omega_p^2}}[2\varepsilon_2( 
\varepsilon_1 -1)+\varepsilon_2^2H^2-\varepsilon_3^2]
$$
\begin{equation}
+{2\omega\omega_H\over{\chi}^2{\omega_p^2} H}\varepsilon_3(\varepsilon_1-1),
\label{3.29}
\end{equation}
\begin{equation}
{\partial\varepsilon_3\over\partial{\chi}}=2
{{\omega_p^2}-\omega^2\over{\chi}^2{\omega_p^2}}\varepsilon_3(\varepsilon_1 
-1)-{\omega\omega_H\over{\chi}^2{\omega_p^2} H}[(\varepsilon_1- 
1)^2+\varepsilon_3^2H^2],
\label{3.30}
\end{equation}
\begin{equation}
{\partial\varepsilon_1\over\partial{\omega_p^2}}=
-{\omega^2\over{\chi}\omega_p^4}[(\varepsilon_1-1)^2+\varepsilon_3^2H^2]
-{2\omega\omega_H H\over{\chi}\omega_p^4}\varepsilon_3(\varepsilon_1-1),
\label{3.31}
\end{equation}
$$
{\partial\varepsilon_2\over\partial{\omega_p^2}}=
-{\omega^2\over{\chi}\omega_p^4}[2\varepsilon_2(\varepsilon_1- 
1)+\varepsilon_2^2H^2-\varepsilon_3^2]
$$
\begin{equation}
+{2\omega\omega_H \over{\chi}\omega_p^4 H}\varepsilon_3(\varepsilon_1-1),
\label{3.32}
\end{equation}
\begin{equation}
{\partial\varepsilon_3\over\partial{\omega_p^2}}=
-{2\omega^2\over{\chi}\omega_p^4}\varepsilon_3(\varepsilon_1-1)
-{\omega\omega_H \over{\chi}\omega_p^4 H}[(\varepsilon_1- 
1)^2+\varepsilon_3^2H^2],
\label{3.33}
\end{equation}
\begin{equation}
{\partial\varepsilon_1\over\partial\omega_H}=
{2\omega H\over{\chi}{\omega_p^2}}\varepsilon_3(\varepsilon_1-1),
\label{3.34}
\end{equation}
\begin{equation}
{\partial\varepsilon_2\over\partial\omega_H}=
-{2\omega \over{\chi}{\omega_p^2} H}\varepsilon_3(\varepsilon_1-1),
\label{3.35}
\end{equation}
\begin{equation}
{\partial\varepsilon_3\over\partial\omega_H}=
{\omega \over{\chi}{\omega_p^2} H}[(\varepsilon_1-1)^2+\varepsilon_3^2H^2].
\label{3.36}
\end{equation}
Using (\ref{3.18}) and the fact that ${\cal D}_i= \varepsilon_{im}{\cal 
E}_m$, we can also write
the tensor (\ref{3.27}) into the form
$$
\langle \Pi_{ij}\rangle =\Big\{
p_0+{H^2\over2}
-
{{\cal E}_l^*{\cal E}_m\over4}\left(\rho 
{\partial\varepsilon_{lm}\over\partial\rho}
-\varepsilon_{lm}\right)
\Big\}\delta_{ij}
-H_iH_j
$$
$$
-{1\over4}
\Big\{
\varepsilon_1{\cal E}_i{\cal E}_j^*+ \varepsilon_2({\bf H}{\cal E}^*){\cal 
E}_iH_j
+\varepsilon_2({\bf H}{\cal E}){\cal E}_i^*H_j
$$
$$
+H_iH_j\Big[
{\partial\varepsilon_1\over\partial H^2}{\cal E}{\cal E}^*+
{\partial\varepsilon_2\over\partial H^2}({\bf H}{\cal E})({\bf H}{\cal E}^*)
+i{\partial\varepsilon_3\over\partial H^2}({\cal E}^* \times{\cal E}){\bf 
H}\Big]
$$
$$
+{i\varepsilon_3\over2}[{\cal E}_i^*({\cal E}\times{\bf H})_j
+{\cal E}_i({\bf H}\times{\cal E}^*)_j+H_i({\cal E}^*\times{\cal E})_j]
$$
\begin{equation}
+(i\leftrightarrow j)
\Big\}.
\label{3.37}
\end{equation}
If the following identity is noted
$$
{\cal E}_i^*({\cal E}\times{\bf H})_j+{\cal E}_i({\bf H}\times{\cal 
E}^*)_j+H_i({\cal E}^*\times{\cal E})_j
$$
\begin{equation}
=({\cal E}^*\times{\cal E}){\bf H}\delta_{ij},
\label{3.38}
\end{equation}
we finally obtain 
$$
\langle \Pi_{ij}\rangle =\Big\{
p_0+{H^2\over2}
-
{{\cal E}_l^*{\cal 
E}_m\over4}\left(\rho{\partial\varepsilon_{lm}\over\partial\rho}-
{\varepsilon_{lm}+\varepsilon_{lm}^*\over2}\right)
\Big\}\delta_{ij}
$$
$$
-{1\over4}
[
\varepsilon_1{\cal E}_i{\cal E}_j^*+\varepsilon_2({\bf H}{\cal E}^*)({\cal 
E}_iH_j+{\cal E}_jH_i)+c.c.]
-H_iH_j
$$
$$
-{H_iH_j\over2}\Big[
{\partial\varepsilon_1\over\partial H^2}{\cal E}{\cal E}^*+
{\partial\varepsilon_2\over\partial H^2}({\bf H}{\cal E})({\bf H}{\cal E}^*)
$$
\begin{equation}
+
i{\partial\varepsilon_3\over\partial H^2}({\cal E}^*\times{\cal E}){\bf 
H}\Big].
\label{3.39}
\end{equation}
This agrees with the Pitaevski${\breve{\rm i}}$'s stress tensor of a 
variable electric
field in a liquid located in a strong magnetic field \cite{pita}.

\subsection{Time-dependent permittivity\label{time}}

In the previous subsections, we assumed that the parameters 
${\chi},~{\omega_p^2}$, $\xi$ and the
magnetic field {\bf B}~ are time-independent. Consequently, the 
permittivity discussed in the subsection
\ref{perm} is static. Now, we will abandon the restriction and
allow ${\chi},{\omega_p^2},\xi,{\bf B}$ to change slowly with time. This 
case is naturally accounted for by (\ref{2.7}), (\ref{2.18}) and 
(\ref{2.20}), the equation of motion for {\bf P}, $$
{\partial\over\partial t}\left({{\dot{\bf P}}\over{\chi}{\omega_p^2}}
+{2\xi{\bf B}\over{\chi}{\omega_p^2}}\times{{\bf P}}\right)
-\left({\partial\over\partial t}{\xi{\bf 
B}\over{\chi}{\omega_p^2}}\right)\times{\bf P}
$$
\begin{equation}
+\left({1\over{\chi}}-1\right){\bf P}-{\bf E}=0.
\label{3.40}
\end{equation}
Here, we again neglected the dissipation ${\bf h}^D$ and considered a 
stationary medium:
$\mbox{\boldmath$v$}({\bf r}, t)\equiv 0$. Comparing the equation 
(\ref{3.40}) with (\ref{3.3}), we can see that the
temporal variations of ${\chi},{\omega_p^2},\xi,{\bf B}$ give rise to 
additional terms,
which result in a dynamic
correction $\varepsilon_{ij}^{\rm dyn}$ to the static dielectric 
permittivity
obtained in subsection III.A. In other words, the relationship between the 
amplitude
of polarization ${\cal P}$ and that of electric field ${\cal E}$ is no 
longer given by (\ref{3.5}), but by
\begin{equation}
{\cal P}_i=(\varepsilon_{ij}-\delta_{ij}){\cal E}_j+\varepsilon_{ij}^{\rm 
dyn}{\cal E}_j,
\label{3.41}
\end{equation}
where $\varepsilon_{ij}$ is given 
by (\ref{3.10}) and (\ref{3.6}-\ref{3.8}).
$\varepsilon_{ij}^{\rm dyn}$ may be calculated by
retaining a monochromatic electric field  in (\ref{3.40}): ${\dot{\cal 
E}}=0$.
Yet, because $\varepsilon$ is now time-dependent,
the amplitude of polarization ${\cal P}$ will change with time, 
${\dot{\cal P}}_i={\dot\varepsilon}_{ij}{\cal E}_j$. And Eq(\ref{3.40}) 
becomes,

$$
\left(1-{1\over{\chi}}+{\omega^2\over{\chi}{\omega_p^2}}\right){\cal P}
+2i{\xi\omega\over{\chi}{\omega_p^2}}{\bf B}\times{\cal P}
$$
$$
+{\cal E}+{\partial\over\partial 
t}\left({2i\omega\over{\chi}{\omega_p^2}}{\hat\varepsilon}
{\cal E} -{2\xi{\bf B}\over{\chi}{\omega_p^2}}\times{\hat\varepsilon}{\cal 
E}\right)
$$
\begin{equation}
-i\omega\left({\partial\over\partial t}{1\over{\chi}{\omega_p^2}}\right)
({\hat\varepsilon}+1){\cal E}
+\left({\partial\over\partial t}{\xi{\bf B}\over{\chi}{\omega_p^2}}\right) 
\times({\hat\varepsilon}+1){\cal E}
=0,
\label{3.42}
\end{equation}
where ${\hat\varepsilon}$ is matrix notation of the permittivity 
(\ref{3.10}).
Solving for ${\cal P}$, we obtain 
$$
\varepsilon_{ij}^{\rm dyn}=(\varepsilon_{im}-\delta_{im})
\Big[
{2i\omega\over{\chi}{\omega_p^2}}{\dot\varepsilon}_{mj}
+i\omega\left({\partial\over\partial t}{1\over{\chi}{\omega_p^2}}\right)
(\varepsilon_{mj}-\delta_{mj})
$$
\begin{equation}
-{2\xi\over{\chi}{\omega_p^2}}\epsilon_{mnl}B_n{\dot\varepsilon}_{lj}
-\epsilon_{mnl}\left({\partial\over\partial t}{\xi 
B_n\over{\chi}{\omega_p^2}}\right) (\varepsilon_{lj}-\delta_{lj})
\Big].
\label{3.43}
\end{equation}
 Together with the static permittivity derived before, this dynamic 
correction provides the expression for the full time-dependent permittivity.

To compare with \cite{pita}, we decompose the dynamic correction
(\ref{3.43}) into a hermitian and an antihermitian part:
\begin{equation}
\varepsilon_{ij}^{\rm dyn}={^\prime\varepsilon_{ij}^{\rm dyn}}
+i~{^{\rm \prime\prime}\varepsilon_{ij}^{\rm dyn}}.
\label{3.44}
\end{equation}
Both matrices ${'\varepsilon^{\rm dyn}}$ and ${''\varepsilon^{\rm dyn}}$
may have complex elements, but must be hermitian.
In accordance to (\ref{3.44}), we may also call them the real and imaginary
parts of $\varepsilon^{\rm dyn}_{ij}$.
Particularly, the imaginary part is
$$
{^{\rm \prime\prime}\varepsilon_{ij}^{\rm dyn}}
={i\over2}(\varepsilon_{ji}^{\rm dyn*}-\varepsilon_{ij}^{\rm dyn}),
$$
which can be also written as
\begin{equation}
{\partial\over\partial t}
\left[
(\varepsilon_{im}-\delta_{im})
\left({\omega\delta_{mn}+i\xi\epsilon_{mkn}B_k\over{\chi}{\omega_p^2}}
\right)(\varepsilon_{nj}-\delta_{nj})
\right],
\label{3.45}
\end{equation}
here the fact $\varepsilon_{ij}^*=\varepsilon_{ji}$ is used.
From the expression (\ref{3.10}) and
 (\ref{3.6}-\ref{3.8})
for $\varepsilon$, one can show that the equation
\begin{equation}
(\varepsilon_{im}-\delta_{im})
\left({\omega\delta_{mn}+i\xi\epsilon_{mkn}B_k\over{\chi}{\omega_p^2}}
\right) (\varepsilon_{nj}-\delta_{nj})
={1\over2}{\partial\varepsilon_{ij}\over\partial\omega}
\label{3.46}
\end{equation}
is valid. So the imaginary part (\ref{3.45}) of the dynamic contribution
to the permittivity is 

\begin{equation}
{^{\rm \prime\prime}\varepsilon_{ij}^{\rm dyn}}
= {1\over2}{\partial^2\varepsilon_{ij}\over\partial\omega\partial t}.
\label{3.47}
\end{equation}
This formula was first obtained in \cite{pita}.

\section{Discussions}\label{dis}

Because both the dispersion and nonlinearity are accounted for,
the hydrodynamic theory of dispersion sketched in section II is a fairly 
complete theory for the dynamics of a fluid interacting with varying fields. 
The theory is derived by generalizing the hydrodynamic approach, but the 
result is consistent with the work of Pitaevski${\breve{\rm i}}$, who starts 
from rather different physics. 

Though not shown here, the present theory reduces to the hydrodynamic 
one~\cite{nurmax},
in the low frequency limit $\omega\to 0$. It is also in agreement
with the Barash-Karpman's stress tensor derived for quasi-monochromatic
field (ie including the lowest-order effects of temporal variation of
the field amplitude ${\cal E}$)\cite{baka}. All these features support
the statement that the basic equations shown in section II are
correctly formulated, particularly the fundamental differential relation 
(\ref{2.2}). 

In our theory for dispersive media, the Pitaevski${\breve{\rm i}}$'s
magnetization appears as a consequence of circular motions 
of the polarization. In contrast to  the conventional magnetization of 
atomic origin, the Pitaevski${\breve{\rm i}}$'s magnetization
is macroscopic. Because the circular motion of ${\bf P}$
is usually accompanied by a rotating electric field, the
Pitaevski${\breve{\rm i}}$'s magnetization is less suitably generated by 
linearly polarized electromagnetic fields, as suggested in \cite{pita}. 
It is remarkable that the explanation to this phenomenon was given 
by the first nonlinear term in the expansion of the energy, Eq(\ref{2.13}). 
We expect it to be an important nonlinear effect, and to play a significant 
role in the nonlinear interaction between matter and intense laser lights.

One author (Y. J.) acknowledges the financial support of SRF for ROCS, SEDC.

\end{multicols}

\begin{thebibliography}{99}
\bibitem{patch} Because the imaginary parts of the permeabilities, 
$\varepsilon''$ and $\mu''$, are odd functions of the frequency $\omega$, 
and the real parts, $\varepsilon'$ and $\mu'$ even ones, and because the 
first frequency dependent effects when leaving the equilibrium are linear in 
$\omega$, they belong to $\varepsilon'', \mu''$ and are dissipative. Lack of 
dissipation therefore characterizes a frequency region disjunct from 
$\omega=0$.
\end{thebibliography}
\end{document}